\documentstyle[twoside]{article}

\catcode`\@=11
\long\def\@makefntext#1{
\protect\noindent \hbox to 3.2pt {\hskip-.9pt  
$^{{\eightrm\@thefnmark}}$\hfil}#1\hfill}		

\def\thefootnote{\fnsymbol{footnote}}
\def\@makefnmark{\hbox to 0pt{$^{\@thefnmark}$\hss}}	
	
\def\ps@myheadings{\let\@mkboth\@gobbletwo
\def\@oddhead{\hbox{}
\rightmark\hfil\eightrm\thepage}   
\def\@oddfoot{}\def\@evenhead{\eightrm\thepage\hfil
\leftmark\hbox{}}\def\@evenfoot{}
\def\sectionmark##1{}\def\subsectionmark##1{}}

\oddsidemargin=\evensidemargin
\renewcommand{\thefootnote}{\fnsymbol{footnote}}

\newcounter{sectionc}\newcounter{subsectionc}\newcounter{subsubsectionc}
\renewcommand{\section}[1] {\vspace{12pt}\addtocounter{sectionc}{1} 
\setcounter{subsectionc}{0}\setcounter{subsubsectionc}{0}\noindent 
	{\tenbf\thesectionc. #1}\par\vspace{5pt}}
\renewcommand{\subsection}[1] {\vspace{12pt}\addtocounter{subsectionc}{1} 
	\setcounter{subsubsectionc}{0}\noindent 
	{\bf\thesectionc.\thesubsectionc. {\kern1pt \bfit #1}}\par\vspace{5pt}}
\renewcommand{\subsubsection}[1] {\vspace{12pt}\addtocounter{subsubsectionc}{1}
	\noindent{\tenrm\thesectionc.\thesubsectionc.\thesubsubsectionc.
	{\kern1pt \tenit #1}}\par\vspace{5pt}}
\newcommand{\nonumsection}[1] {\vspace{12pt}\noindent{\tenbf #1}
	\par\vspace{5pt}}

\newcounter{appendixc}
\newcounter{subappendixc}[appendixc]
\newcounter{subsubappendixc}[subappendixc]
\renewcommand{\thesubappendixc}{\Alph{appendixc}.\arabic{subappendixc}}
\renewcommand{\thesubsubappendixc}
	{\Alph{appendixc}.\arabic{subappendixc}.\arabic{subsubappendixc}}

\renewcommand{\appendix}[1] {\vspace{12pt}
        \refstepcounter{appendixc}
        \setcounter{figure}{0}
        \setcounter{table}{0}
        \setcounter{lemma}{0}
        \setcounter{theorem}{0}
        \setcounter{corollary}{0}
        \setcounter{definition}{0}
        \setcounter{equation}{0}
        \renewcommand{\thefigure}{\Alph{appendixc}.\arabic{figure}}
        \renewcommand{\thetable}{\Alph{appendixc}.\arabic{table}}
        \renewcommand{\theappendixc}{\Alph{appendixc}}
        \renewcommand{\thelemma}{\Alph{appendixc}.\arabic{lemma}}
        \renewcommand{\thetheorem}{\Alph{appendixc}.\arabic{theorem}}
        \renewcommand{\thedefinition}{\Alph{appendixc}.\arabic{definition}}
        \renewcommand{\thecorollary}{\Alph{appendixc}.\arabic{corollary}}
        \renewcommand{\theequation}{\Alph{appendixc}.\arabic{equation}}
        \noindent{\tenbf Appendix \theappendixc #1}\par\vspace{5pt}}
\newcommand{\subappendix}[1] {\vspace{12pt}
        \refstepcounter{subappendixc}
        \noindent{\bf Appendix \thesubappendixc. {\kern1pt \bfit #1}}
	\par\vspace{5pt}}
\newcommand{\subsubappendix}[1] {\vspace{12pt}
        \refstepcounter{subsubappendixc}
        \noindent{\rm Appendix \thesubsubappendixc. {\kern1pt \tenit #1}}
	\par\vspace{5pt}}

\newcommand{\textlineskip}{\baselineskip=13pt}
\newcommand{\smalllineskip}{\baselineskip=10pt}

\def\eightcirc{
\begin{picture}(0,0)
\put(4.4,1.8){\circle{6.5}}
\end{picture}}
\def\eightcopyright{\eightcirc\kern2.7pt\hbox{\eightrm c}} 

\newcommand{\copyrightheading}[1]
	{\vspace*{-2.5cm}\smalllineskip{\flushleft
	{\footnotesize International Journal of Modern Physics A, #1}\\
	{\footnotesize $\eightcopyright$\, World Scientific Publishing
	 Company}\\
	 }}

\def\abstracts#1#2#3{{
	\centering{\begin{minipage}{4.5in}\baselineskip=10pt\footnotesize
	\parindent=0pt #1\par 
	\parindent=15pt #2\par
	\parindent=15pt #3
	\end{minipage}}\par}} 

\renewenvironment{thebibliography}[1]
	{\frenchspacing
	 \ninerm\baselineskip=11pt
	 \begin{list}{\arabic{enumi}.}
	{\usecounter{enumi}\setlength{\parsep}{0pt}
	 \setlength{\leftmargin 12.7pt}{\rightmargin 0pt} 
	 \setlength{\itemsep}{0pt} \settowidth
	{\labelwidth}{#1.}\sloppy}}{\end{list}}

\newcounter{itemlistc}
\newcounter{romanlistc}
\newcounter{alphlistc}
\newcounter{arabiclistc}
\newenvironment{itemlist}
    	{\setcounter{itemlistc}{0}
	 \begin{list}{$\bullet$}
	{\usecounter{itemlistc}
	 \setlength{\parsep}{0pt}
	 \setlength{\itemsep}{0pt}}}{\end{list}}

\newcommand{\fcaption}[1]{
        \refstepcounter{figure}
        \setbox\@tempboxa = \hbox{\footnotesize Fig.~\thefigure. #1}
        \ifdim \wd\@tempboxa > 5in
           {\begin{center}
        \parbox{5in}{\footnotesize\smalllineskip Fig.~\thefigure. #1}
            \end{center}}
        \else
             {\begin{center}
             {\footnotesize Fig.~\thefigure. #1}
              \end{center}}
        \fi}

\newcommand{\tcaption}[1]{
        \refstepcounter{table}
        \setbox\@tempboxa = \hbox{\footnotesize Table~\thetable. #1}
        \ifdim \wd\@tempboxa > 5in
           {\begin{center}
        \parbox{5in}{\footnotesize\smalllineskip Table~\thetable. #1}
            \end{center}}
        \else
             {\begin{center}
             {\footnotesize Table~\thetable. #1}
              \end{center}}
        \fi}

\def\@citex[#1]#2{\if@filesw\immediate\write\@auxout
	{\string\citation{#2}}\fi
\def\@citea{}\@cite{\@for\@citeb:=#2\do
	{\@citea\def\@citea{,}\@ifundefined
	{b@\@citeb}{{\bf ?}\@warning
	{Citation `\@citeb' on page \thepage \space undefined}}
	{\csname b@\@citeb\endcsname}}}{#1}}

\newif\if@cghi
\def\cite{\@cghitrue\@ifnextchar [{\@tempswatrue
	\@citex}{\@tempswafalse\@citex[]}}
\def\citelow{\@cghifalse\@ifnextchar [{\@tempswatrue
	\@citex}{\@tempswafalse\@citex[]}}
\def\@cite#1#2{{$\null^{#1}$\if@tempswa\typeout
	{IJCGA warning: optional citation argument 
	ignored: `#2'} \fi}}

\def\pmb#1{\setbox0=\hbox{#1}
	\kern-.025em\copy0\kern-\wd0
	\kern.05em\copy0\kern-\wd0
	\kern-.025em\raise.0433em\box0}

\def\fnt#1#2{\footnotetext{\kern-.3em
	{$^{\mbox{\scriptsize #1}}$}{#2}}}

\def\fpage#1{\begingroup
\voffset=.3in
\thispagestyle{empty}\begin{table}[b]\centerline{\footnotesize #1}
	\end{table}\endgroup}

\def\runninghead#1#2{\pagestyle{myheadings}
\markboth{{\protect\footnotesize\it{\quad #1}}\hfill}
{\hfill{\protect\footnotesize\it{#2\quad}}}}
\font\tenrm=cmr10
\font\tenit=cmti10 
\font\tenbf=cmbx10
\font\bfit=cmbxti10 at 10pt
\font\ninerm=cmr9

\font\eightrm=cmr8

\def\qed{\hbox{${\vcenter{\vbox{			
   \hrule height 0.4pt\hbox{\vrule width 0.4pt height 6pt
   \kern5pt\vrule width 0.4pt}\hrule height 0.4pt}}}$}}

\renewcommand{\thefootnote}{\fnsymbol{footnote}}	

\begin{document}

\runninghead{Light-cone form of field dynamics 
$\ldots$} {Light-cone form of field dynamics
$\ldots$}

\normalsize\textlineskip
\thispagestyle{empty}
\setcounter{page}{1}

\copyrightheading{}			

\begin{flushright}
FIAN/TD/00-17
\\
OHSTPY-HEP-T-00-027
\\
hep-th/0011112\\
\end{flushright}
\vspace{-1.5cm}

\vspace*{0.88truein}

\fpage{1}
\centerline{\bf LIGHT-CONE FORM OF FIELD DYNAMICS }
\vspace*{0.035truein}
\centerline{\bf IN ADS SPACE-TIME}
\vspace*{0.37truein}
\centerline{\footnotesize R.R. METSAEV\footnote{
Permanent address:
Department of Theoretical Physics, P.N. Lebedev Physical
Institute, Leninsky prospect 53,  Moscow 117924, Russia.
E-mail: metsaev@pacific.mps.ohio-state.edu}}
\vspace*{0.015truein}
\centerline{\footnotesize\it 
Department of Physics,
The Ohio State University,}
\baselineskip=10pt
\centerline{\footnotesize\it 
174 West 18th Avenue, Columbus, OH 43210-1106, USA}
\vspace*{10pt}

\vspace*{0.21truein}
\abstracts{
Light-cone approach to field dynamics in AdS space-time is discussed.
}{}{}


\vspace*{1pt}\textlineskip	

\textheight=7.8truein
\setcounter{footnote}{0}
\renewcommand{\thefootnote}{\alph{footnote}}

\section{Introduction}	        
\vspace*{-0.5pt}
\noindent
In spite of its Lorentz noncovariance, the light-cone formalism\cite{1}
offers conceptual and technical simplifications of approaches  to 
various problems of modern quantum field theory. For example,  one can
mention the construction of first quantized light-cone string
action\cite{2} and manifestly supersummetric  formulation for 
superfield theories of superstrings\cite{3,4}. Another attractive
application of the light-cone formalism is a construction of
interaction vertices in the theory of higher spin  massless
fields\cite{5,6,7,8,9}.  Note that sometimes, a theory formulated
within this formalism turns out to be a good starting point for
deriving a Lorentz covariant formulation\cite{10}. 

Motivated by a desire to solve the problem\footnote{Another problem
which triggered our investigation of $AdS$ light-cone  formalism is a
construction of action for the equations of motion of higher spin
massless field  theory\cite{11}.}\ \ of  $AdS$ superstring\cite{12,12.1} a
light-cone  gauge form of field dynamics in $AdS$ space-time was
recently developed\cite{13,14}. Application of light-cone formalism  to
$IIB$ supergravity in $AdS_5\times S^5$ background and discussion  of
various related issues may be found in Refs.\cite{15,16}. Study of
$AdS$ superstring in the framework of light-cone gauge may be found  in
Refs.\cite{17,18}. In the particle theory limit, the string Hamiltonian
\cite{17,18} reduces to the light-cone Hamiltonian for a 
superparticle\cite{15} in $AdS_5\times S^5$.  This {\it implies}  that
the ``massless" (zero-mode)  spectrum of the superstring light-cone
gauge action\cite{17,18} coincides indeed with the spectrum of type IIB
supergravity compactified on $S^5$. Discussion of alternative gauges
for the string in  $AdS_5\times S^5$ background may be found in 
\cite{19,20,21,22,23}. Here we restrict our attention to light-cone
gauge field dynamics in $AdS$ space-time.


\section{Light-cone gauge action}
\noindent

Let $\phi$ be arbitrary spin field propagating in $AdS$ space-time. 
Light-cone gauge action for this field can be cast into the following 
`nonrelativistic form'\cite{13}\footnote{
We use parametrization of $AdS_d$ space in which
$ds^2=R^2(-dx_0^2+dx_I^2+dx_{d-1}^2)/z^2$, where $R$ is radius of $AdS$
geometry.
Light-cone
coordinates in $\pm$ directions are defined as $x^\pm=(x^{d-1}\pm
x^0)/\sqrt{2}$ and we adopt the following  conventions:
$I,J=1,\ldots, d-2$; $i,j,k,l=1,\ldots,d-3$.
$\partial^I=\partial_I\equiv\partial/\partial x^I$,
$\partial^\pm=\partial_\mp \equiv \partial/\partial x^\mp$,
$z\equiv x^{d-2}$. The coordinate $x^+$ is taken to be a light-cone time.}

\begin{equation}\label{lcaction}
S_{l.c.}
= \int d^dx  \partial^+\phi(-\partial^- +P^-)\phi\,,
\qquad
P^-=-\frac{\partial_I^2}{2\partial^+}
+\frac{1}{2z^2\partial^+}A\,,
\end{equation}
where $P^-$ is the (minus) Hamiltonian density  and $A$ is some
operator does not depending on space-time coordinates and  their
derivatives. This operator acts only on spin indices of field $\phi$.
From the expressions above it is clear that the action can be rewritten
in the following `covariant form'

\begin{equation}\label{2lcact}
S_{l.c.}
=\frac{1}{2}\int d^dx \phi\bigl(\Box
-\frac{1}{z^2}A\bigr)\phi\,,
\qquad
\Box = 2\partial^+\partial^- + \partial^I\partial^I\,,
\end{equation}
where $\Box$ is the flat D'Alembertian operator.
We shall call the operator $A$ the $AdS$ mass operator\footnote{
Note that the $AdS$ mass operator $A$ for massless
fields does not equal to zero in general. 
The operator $A$ is equal
to zero only for massless representations which can be realized as
irreducible representations of conformal algebra\cite{24} which for
the case of $d$-dimensional AdS space-time is the $so(d,2)$ algebra.}.
By now this operator is known for the following cases
\begin{itemlist}
 \item massive fields of arbitrary spin and arbitrary type of 
 Young symmetry\cite{13}
 \item massless fields of arbitrary spin corresponding to 
 totally symmetric and
 totally antisymmetric representations of $so(d-2)$ algebra\cite{13}
\item type IIB supergravity in $AdS_5 \times S^5$ background\cite{15}
\end{itemlist}
Ror the case of massless mixed symmetry fields
the $AdS$ mass operator is still to be found\footnote{
The gauge invariant equations of motion for such fields 
taken to be in Lorentz gauge
have been found in Refs\cite{25,26}.
Because in light-cone gauge the Lorentz constraint does not follow
from the AdS gauge invariant equations of motion we cannot use these 
equations to develop light-cone formulation.
Recently for the case of mixed symmetry fields 
it was demonstrated\cite{27} that in contrast to massless fields in
Minkowski space whose physical degrees of freedom transform in irreps
of $so(d-2)$ algebra, $AdS$ massless mixed symmetry fields reduce to a
number of irreps of $so(d-2)$ algebra.}.
Let us discuss $AdS$ mass operator for various fields.

{\it Massive scalar field.} In this case starting with the standard 
action

\begin{equation}
S=\frac{1}{2}\int d^dx \sqrt{g}
\Phi(\frac{1}{\sqrt{g}}\partial_\mu \sqrt{g}g^{\mu\nu}\partial_\nu
-m^2)\Phi
\end{equation}
and making rescaling $\Phi=z^{(d-2)/2}\phi$ one finds an action similar to 
(\ref{2lcact}) where operator $A$ takes the form

\begin{equation}
A=(mR)^2+\frac{d(d-2)}{4}\,.
\end{equation}
For conformal invariant scalar field one has
$m^2=-\frac{d(d-2)}{4R^2}$ and this gives $A=0$.

{\it Massless spin 1 field.} In light-cone gauge physical d.o.f. of
spin one  massless field are described by vector field $\phi^I$. 
It turns out
that the field $\phi^I$ is not eigenvector of the operator $A$.
However if  we decompose the field $\phi^I$, which is $so(d-2)$
vector,  into $so(d-3)$ vector $\phi_1^i\equiv \phi^i$  and $so(d-3)$
scalar  $\phi_0\equiv \phi^z$ then one has\footnote{ Note that all
tensor fields are defined in tangent space\cite{13}.}

\begin{equation}
A\phi_1^i=\frac{(d-2)(d-4)}{4}\phi_1^i\,,
\qquad
A\phi_0=\frac{(d-4)(d-6)}{4}\phi_0\,,
\end{equation}
and the light-cone gauge action (\ref{2lcact}) takes the form

\begin{equation}
S_{l.c.}
=\frac{1}{2}\int d^dx  \Bigl(\phi^I\Box\phi^I
-\frac{(d-2)(d-4)}{4z^2}|\phi_1|^2
-\frac{(d-4)(d-6)}{4z^2}|\phi_0|^2\Bigr)\,.
\end{equation}

{\it Massless spin 2 field - graviton.} 
Physical d.o.f. of spin two field are described by symmetric tracesless
$so(d-2)$ tensor field $\phi^{IJ}$. As before this field is not
eigenvector of $A$. Decomposing $\phi^{IJ}$ into $so(d-3)$
traceless tensor field $\phi_2^{ij}\equiv
\phi^{ij}-\delta^{ij}\phi^{kk}/(d-3)$, vector field  $\phi_1^i\equiv
\phi^{zi}$ and scalar field $\phi_0\equiv \phi^{zz}$  we get

\begin{equation}
A\phi_2^{ij}=\frac{d(d-2)}{4}\phi_2^{ij}\,,
\qquad
A\phi_1^i=\frac{(d-2)(d-4)}{4}\phi_1^i\,,
\qquad
A\phi_0=\frac{(d-4)(d-6)}{4}\phi_0\,,
\end{equation}
and the following light-cone gauge action

\begin{equation}
S_{l.c.}
=\frac{1}{2}\!\int\!\! d^dx(\phi^{IJ}\Box\phi^{IJ}
-\frac{d(d-2)}{4z^2}|\phi_2|^2
-\frac{(d-2)(d-4)}{4z^2}|\phi_1|^2
-\frac{(d-4)(d-6)}{4z^2}|\phi_0|^2)
\end{equation}

{\it Massless arbitrary spin totally symmetric field.} 
In this case we start with totally 
symmetric tracesless $so(d-2)$ tensor field $\phi^{I_1\ldots I_s}$
and decompose it into irreducible representations of
$so(d-3)$ algebra $\phi^{I_1\ldots I_s}
=\sum_{s^\prime =0}^s \phi_{s^\prime}^{i_1\ldots i_{s^\prime}}$.
Taking into account

\begin{equation}
A\phi_{s^\prime}^{i_1\ldots i_{s^\prime}}
=A_{s^\prime}\phi_{s^\prime}^{i_1\ldots i_{s^\prime}}\,,
\qquad
A_{s^\prime}
=\Bigl(s^\prime +\frac{d-5}{2}\Bigr)^2-\frac{1}{4}\,,
\end{equation}
we get

\begin{equation}
S_{l.c.}
=\frac{1}{2}\int d^dx\sum_{s^\prime=0}^s(
\phi_{s^\prime}^{i_1\ldots i_{s^\prime}}\Box 
\phi_{s^\prime}^{i_1\ldots i_{s^\prime}}
-\frac{1}{z^2}A_{s^\prime}|\phi_{s^\prime}|^2)\,.
\end{equation}
Note that the above formulas for $AdS$ mass operator are valid for $d>4$. 
The representation for operator $A$ which is valid also for the case of $d=4$ 
is given by\cite{13}

\begin{equation}
A= -\frac{1}{2}M^{ij}M^{ij} +\frac{(d-4)(d-6)}{4}\,,
\qquad
M^{ij}=-M^{ji}\,,
\end{equation}
where $M^{ij}$ is a spin part of $so(d-3)$ angular momentum. 
From these
formulas  we see that for $d=4$ the $AdS$ mass operator is equal to
zero. This reflects the fact that all massless fields in $d=4$ can be
realized as irreducible representations of conformal algebra 
$so(4,2)$.

\nonumsection{Acknowledgements}
\noindent

This  work  was  supported  by
the DOE grant DE-FG02-91ER-40690, by the INTAS project 991590,
and by the RFBR Grant No.99-02-16207.

\nonumsection{References}

\end{document}